# 2D Black Phosphorus Carbide: Rippling and Formation of Nanotubes

*Stepan A. Shcherbinin†, Kun Zhou‡, Sergey V. Dmitriev§, ‖, Elena A. Korznikova§, Artur R. Davletshin⊥, \*Andrey A. Kistanov‡*

†Southern Federal University, 344006 Rostov-on-Don, Russia

‡School of Mechanical and Aerospace Engineering, Nanyang Technological University, 639798 Singapore

§Institute for Metals Superplasticity Problems, Russian Academy of Science, 450001 Ufa, Russia

‖National Research Tomsk State University, 634050 Tomsk, Russia

⊥Department of Petroleum and Geosystems Engineering, The University of Texas at Austin, Austin 78712, TX, USA

‡Nano and Molecular Systems Research Unit, University of Oulu, 90014 Oulu, Finland

**ABSTRACT:** The allotropes of a new layered material, phosphorus carbide (PC), have been predicted recently and a few of these predicted structures have already been successfully fabricated. Herein, by using first-principles calculations we investigated the effects of rippling an α-PC monolayer, one of the most stable modifications of layered PC, under large compressive strains. Similar to phosphorene, layered PC was found to have the extraordinary ability to bend and form ripples with large curvatures under a sufficiently large strain applied along its armchair direction. The band gap size, workfunction, and Young's modulus of rippled α-PC monolayer are predicted to be highly tunable by strain engineering. Moreover, a direct-indirect band gap transition is observed under the compressive strains in a range from 6 to 11%. Another important feature of α-PC monolayer rippled along the armchair direction is the possibility of its rolling to a PC nanotube (PCNT) under extreme compressive strains. These tubes of different sizes exhibit high thermal stability, possess a comparably high Young's modulus, and a well tunable band gap which can vary from 0 to 0.95 eV. In addition, for both structures, rippled α-PC and PCNTs, we have explained the changes of their properties under compressive strain in terms of the modification of their structural parameters.

## 1. INTRODUCTION

The modern experimental techniques developed in the last decade, such as molecular beam epitaxy,[1,2] atomic layer deposition,[3,4] pulsed laser deposition,[5,6] and magnetron sputtering[7] have contributed to a giant leap in the synthesis of two-dimensional (2D) materials, as well as the production of technological devices based on these materials.[8-11] For example, the forefather of 2D materials, graphene, was obtained using mechanical exfoliation.[1,12] Another classical 2D material, phosphorene, was fabricated via liquid-phase exfoliation[13] and pulsed laser exfoliation.[5] One should not underestimate the contribution of atomistic simulations in the investigations of 2D materials. For instance, first-principles calculations predicted the existence of group V elements, arsenene and antimonene,[14,15] for the first time.

The above-mentioned technologies have rendered 2D materials no longer a vision from the future, but materials of present-day and have promoted an extensive study of their structure, properties, and applications. However, contemporary knowledge on 2D materials concludes that each 2D monolayer possesses some inherent disadvantageous properties. Graphene in spite of its unique chemical stability, high mechanical durability, and excellent carrier mobility, has no intrinsic band gap.[16-20] Transition metal dichalcogenides[21,22] possess sizable and tunable band gaps, but they present several disadvantages, depending on their fabrication methods, such as a relatively low carrier mobility, larger volume expansion and phase conversion.[23] A recently predicted and isolated group of materials such as phosphorene, arsenene, and antimonene have extraordinary properties,[14, 24-26] including remarkable charge mobility and wide and highly tunable band gap (0.3–2 eV). However, these materials are chemically active and structurally unstable in standard ambient conditions.[27-32]

For this reason, a new class of materials, the hybridized composites that combine the advantages and counteract the disadvantages of its constituent compounds are currently in the spotlight.[33-38] These materials can exist as heterostructures consisting of several layered 2D materials[33,39-41] such as graphene/phosphorene, graphene/InSe, and boron nitride/phosphorene or 2D layers consisting of atoms of different elements.[42-44] It has been demonstrated that heterostructures formed by 2D materials, which are unstable in strand ambient conditions, such as phosphorene, and 2D materials, which are stable, such as graphene, are characterized by improved stabilities.[37,38,45,46] Very recently, a stable allotrope of phosphorus carbide (PC), a 2D monolayer consisting of carbon and phosphorus atoms, has been discovered.[47-50] It has a wide band gap reaching up to 2.65 eV,[48] robust superconducting behaviour even under a tensile strain,[51] a small effective mass, and extra high carrier mobility.[52]

It is well known that the different properties of 2D materials can be modified by strain engineering.[53-57] For phosphorene, it has been shown that in-plane strains cause its band gap transitions and affects its charge carrier mobility.[56, 57] The strong spatial dependence of the electronic structure upon the rippling of phosphorene along its periodic line profile, which may be used for modulating the injection and confinement of its charge carriers, has been established both theoretically[58] and experimentally.[59] However, there have been no similar systematic studies conducted on the rippling of PC monolayer under a significant out-of-plane deformation.

Graphene and phosphorene nanotubes are currently the subjects of intensive research as base materials for a large variety of technological applications from sensing devices to active optoelectronic elements.[60] For example, phosphorene nanotubes are direct band gap semiconductors and their electronic properties can be modified by strain or electric field engineering.[61] Meanwhile, graphene nanotubes are common additives to polymer hosts for the manufacturing of advanced composite materials[62,63] and as electrode materials for batteries and supercapacitors.[64] However, PC nanotubes (PCNTs) are much less studied.

In this work, we systematically investigated the evolution of the atomic structure, electronic and mechanical properties of rippled α-PC monolayer under large compressive strains. Based on the work,[65] where layered PC, similar to black phosphorene, was found to be soft when strained along its armchair direction compared with a zigzag direction, we focused on the armchair rippled α-PC monolayer due to its high flexibility. After discovering that α-PC monolayer is able to form nanotubes under large compressive strains applied along its armchair direction, we have also investigated PCNTs. In particular, their thermal stability, atomic structure and electronic and mechanical properties were analyzed, which to the best of our knowledge, has not been conducted.

## 2. COMPUTATIONAL METHODS

This work is based on the spin-polarized density-functional theory (DFT) as implemented in the VASP code.[66] The Perdew−Burke−Ernzerhof (PBE) exchange-correlation functional under the generalized gradient approximation (GGA)[67] were used for the geometry optimization and electronic structure calculations. Since in the GGA PBE approach the fundamental band gap is usually underestimated, the HSE06[68] hybrid exchange-correlation functional was adopted to obtain more accurate band structures. The cutoff energy for plane waves was set at 450 eV. The geometries of all structures were fully optimized until the total energy and all forces on atoms converge to less than

$10^{-8}$ eV and 0.01 eV/Å, respectively. The vacuum depth of 15 Å was introduced for all the structures, for both monolayers and nanotubes, to avoid artificial interaction in a supercell. The periodic boundary conditions were applied for the two in-plane transverse directions. The considered supercell of α-PC monolayer was composed of 3×4 unit cells (12 C and 12 P atoms). The study of α-PC nanotubes (PCNTs) commenced with nanotubes of the smallest size possible (4 C atoms and 4 P atoms). In this work, a PCNT consisting of 40 atoms (PCNT40) was considered, to compromise for the computational demand.

The rippled structures of α-PC monolayer were created by applying a compressive strain along the armchair directions. The compressive strain was defined as

$$\varepsilon = \frac{l - l_0}{l_0} \quad (1)$$

where $l$ and $l_0$ are the lattice constants of the strained and initial supercells, respectively. The compression was performed in steps. After each step, the structure was fully relaxed through the technique of energy minimization. Because of the periodic boundary conditions with a fixed period applied laterally, along the armchair direction, the lattice spacing along the zigzag direction remained unchanged.

The Young's modulus of the considered PCNTs was calculated with the aid of the energy-strain relation.[69,70] A set of positive and negative uniaxial strains $\varepsilon$ were applied to the initially unstrained structure along a zigzag direction. For each case considered, the corresponding energy values were calculated. The obtained set of energy values corresponding to each value of the applied strain was then used for a polynomial fitting of energy potential $U(\varepsilon)$. The Young's modulus was computed by

$$E = \frac{1}{V}\left(\frac{\partial^2 U(\varepsilon)}{\partial \varepsilon^2}\right)_{\varepsilon=0}, \quad (2)$$

where $V$ is the volume of the unit cell of the considered PCNT. Here, we assumed $V = 2 \cdot \pi \cdot R \cdot l$, where $R$ is the mean radius of nanotube, while $l$ is the interlayer distance of the α-PC monolayer ($l$ = 4.93 Å, as obtained from DFT calculations). As a consequence of the applied uniaxial tensile strain $\varepsilon$ a negative radial strain $\varepsilon_r$ was naturally obtained. The Poisson's ratio $v$ was calculated by plotting $\varepsilon$ vs. $\varepsilon_r$ and taking the negative of the slope, where

$$v = -\frac{\varepsilon_r}{\varepsilon} \quad (3)$$

The elastic properties and Young's modulus of the α-PC monolayer were calculated similarly to those of rippled structures. *Ab initio* molecular dynamics (AIMD) calculations[71] were conducted at 300 K to verify the thermal stability of PCNTs. The simulation lasted for 10 ps with a time step of 1.0 fs and the temperature was controlled by a Nose–Hoover thermostat.

## 3. RESULTS AND DISCUSSION

**3.1. Electronic structure and mechanical properties of the α-PC monolayer.** We considered the α-PC monolayer since it was found to be one of the stable allotropes of PC.[48,72] Moreover, α-PC monolayer was predicted to possess distinguishable features,[47-49,52] such as a wide band gap and a high Young's modulus. First, the geometry, electronic structure, and Young's modulus of the α-PC monolayer were investigated and the results were compared with those from recent works. The considered α-PC structure and its unit cell (bounded by the dashed line) are presented in Fig. 1a. The obtained relaxed lattice constants $a$ = 8.56 and $b$ = 2.92 Å were in good agreement with the previously reported results.[47-49] Our HSE06 calculations set the band gap size to 1.32 eV. The PBE GGA calculations predicted a qualitatively similar band structure, but slightly underestimated the band gap size, which is 0.76 eV in that case. Both values match the reference results obtained by the HSE06[48] and PBE GGA[48,49] functionals.

To avoid the high computational cost of hybrid functional calculations, we reported the results based on the PBE GGA approach. The Young's moduli of the α-PC monolayer along its armchair and zigzag directions was found to be 27.08 and 348.69 GPa, respectively. The obtained Young's moduli were smaller than those of graphene,[73] but significantly higher than those of phosphorene.[74] Interestingly, the α-PC monolayer was found to possess a mechanical strength similar to a γ-PC monolayer and superior to that of phosphorene.[50] The value of the shear modulus of the α-PC monolayer reached up to 81.80 GPa and the Poisson's ratios are 0.25 for tension along the armchair direction and 0.019 along the zigzag direction.

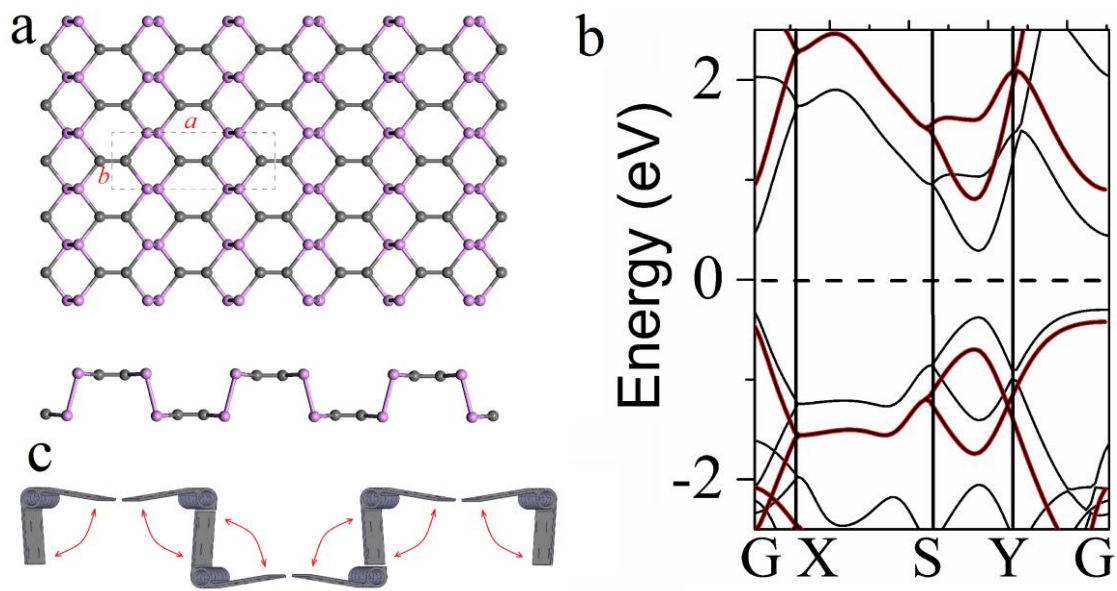

**Figure 1.** (a) The structure of the α-PC monolayer, top and side views. (b) The band structure of α-PC monolayer. The black and red lines show the band structure calculated by GGA and HSE methods, respectively. (c) A schematic hinge-like structure of α-PC monolayer.

**3.2. Atomic structure, electronic and mechanical properties of rippled α-PC.** To examine the extreme effects of strain engineering on the rippling of α-PC monolayer and the possibility of formation of nanotubes as a result of compression, we applied compressive strains along its armchair direction (in which the material is more flexible). Fig. S1 shows the atomic (upper panel) and band (lower panel) structures of rippled α-PC monolayer under compressive strain ranging from 0 to 48%. The correlation between the band gap size and the applied strain is depicted in Fig. 2a. First, for the compressive strain ranging from 0% to 11%, the band gap size slightly decreases from 0.76 to 0.47 eV. At the same time, there are no perceptible changes to the structure of rippled α-PC monolayer within this range of applied strains. Further compression (to 16%) causes a drastic decrease of the band gap size to 0.22 eV. Next, for the compressive strain values of 22 and 28%, the band gap size jumps to 0.54 and 0.56 eV, respectively, and finally decreases linearly to 0.43 eV as the compressive strain reaches ~48%. It should be noted that an indirect-direct band gap transition occurs in α-PC monolayer when the compressive strain is between 6 to 11%. Meanwhile, at the compressive strain higher 11% a direct band gap is observed and the valence band maximum and conduction bands minimum shift from the Γ to Y point.

Since rippling usually changes the workfunction of 2D materials,[75,76] we next examined the workfunction of rippled α-PC monolayer. Fig. 2b shows the workfunction of α-PC monolayer as a function of the applied compressive strain. The workfunction of unstrained monolayered α-PC is 4.77 eV, which is higher than that of graphene (4.50 eV)[77] and lower than that of phosphorene (5.04-5.16 eV).[78] Moreover, the workfunction of the α-PC monolayer was found to have a general tendency to increase with compression strain, but experienced a sharp drop when the strain ranged from 11 to 16%. The found increase of the workfunction of α-PC may be attributed to the role of stress accumulated upon the increase of the strain and stretching of the atomic bonds which leads to the shift of the valence band maximum (see Fig. S1).[79] Similar linear increase of the workfunction caused by in-plane strains has been recently reported for MoS$_2$ nanosheets.[80]

According to Fig. 2c, the Young's modulus of rippled α-PC monolayer significantly decreases from 348.69 to 290.4 GPa as the compressive strain is increased from 0 to 16%. A drastic plummet of the Young's modulus occurs at a compressive strain of 16% at the moment of the major stress accumulation. The subsequent increase of the compressive strain to 32% induces the Young's modulus to rapidly increase up to 341.43 GPa. As the compressive strain increases from 32 to 48%, the Young's modulus slightly decreases to 334.33 GPa.

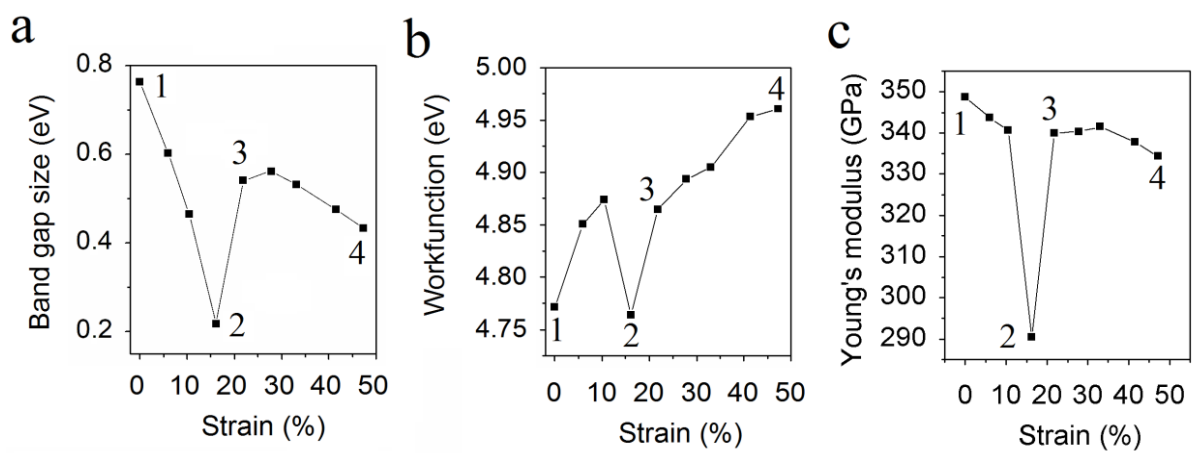

**Figure 2.** (a) The band gap size, (b) workfunction, and (c) Young's modulus of rippled α-PC monolayer as a function of compressive strain. The points marked by numbers 1 to 4 correspond to the structures shown in Fig. 3a.

From the analysis presented, it is evident that there is an accumulation of stress upon increasing the strain to ~11%, while as the compressive stain reaches ~16%, a stress relaxation occurs which results in a decrease of the band gap size, workfunction, and Young's modulus. Further increase in compressive strain, after the release of stress, results in a sharp increase of the band gap size, workfunction, and Young's modulus of rippled α-PC. This can be explained by a specific hinge-like structure of α-PC in its armchair direction (Fig. 1c, lower panel), which allows α-PC to withstand large compressive strains without breaking, similar to phosphorene.[56,58] To demonstrate this feature, the following structural parameters of rippled α-PC monolayer during the compression deformation were tracked: the bond lengths P-P1 and P-P2 connecting the hinges, the bond lengths of the hinges C-C1, C-C2, C-P1, and C-P2, and the hinge angles CPC1, CPC2, PCC1, PCC2, PPC1, and PPC2, as shown in the lower panel of Fig. 3a. The variation of the bond lengths and angles are presented in Figs. 3b and c, respectively.

Indeed, for a compressive strain up to 16%, there are no remarkable changes in the structural parameters except for a slight elongation of the P-P1 and P-P2 bonds and a slight decrease of the PPC1 and PPC2 angles. Determinative changes of the structural parameters begin at a compressive strain of about 16% and higher. The following changes are observed: i) oscillation of P-P1 and P-P2 bond and a slight elongation of C-C1 and C-C2 bonds (Fig. 3b), ii) a decrease of the PCC1 angle, and iii) a decrease of the PPC2 and an increase of the PPC1 angles (Fig. 3c). The observed discrepancies of the angles indicate that the structure folds like a hinge upon reaching a central level of compressive straining (~16%), which facilitates the relaxation of accumulated stress. When the compressive strain exceeds 50%, breaking of two P-P bonds at the base of the structure and the formation of a single P-P bond occurs with the restructuration of rippled α-PC to form a nanotube-like structure. Our AIMD calculations show that transformation of α-PC from the rippled structure into the nanotube-like structure under a compressive strain of ~50% occurs within ~4ps (Movie 1 in SI). The nanotube is unstable in a long period of time because in this case, however, from this simulation, the possibility of rolling of α-PC ripple to nanotube is revealed. Previously, strain induced rolling into nanotube has been predicted experimentally for the $In_xGa_{1-x}As$-GaAs membranes.[81]

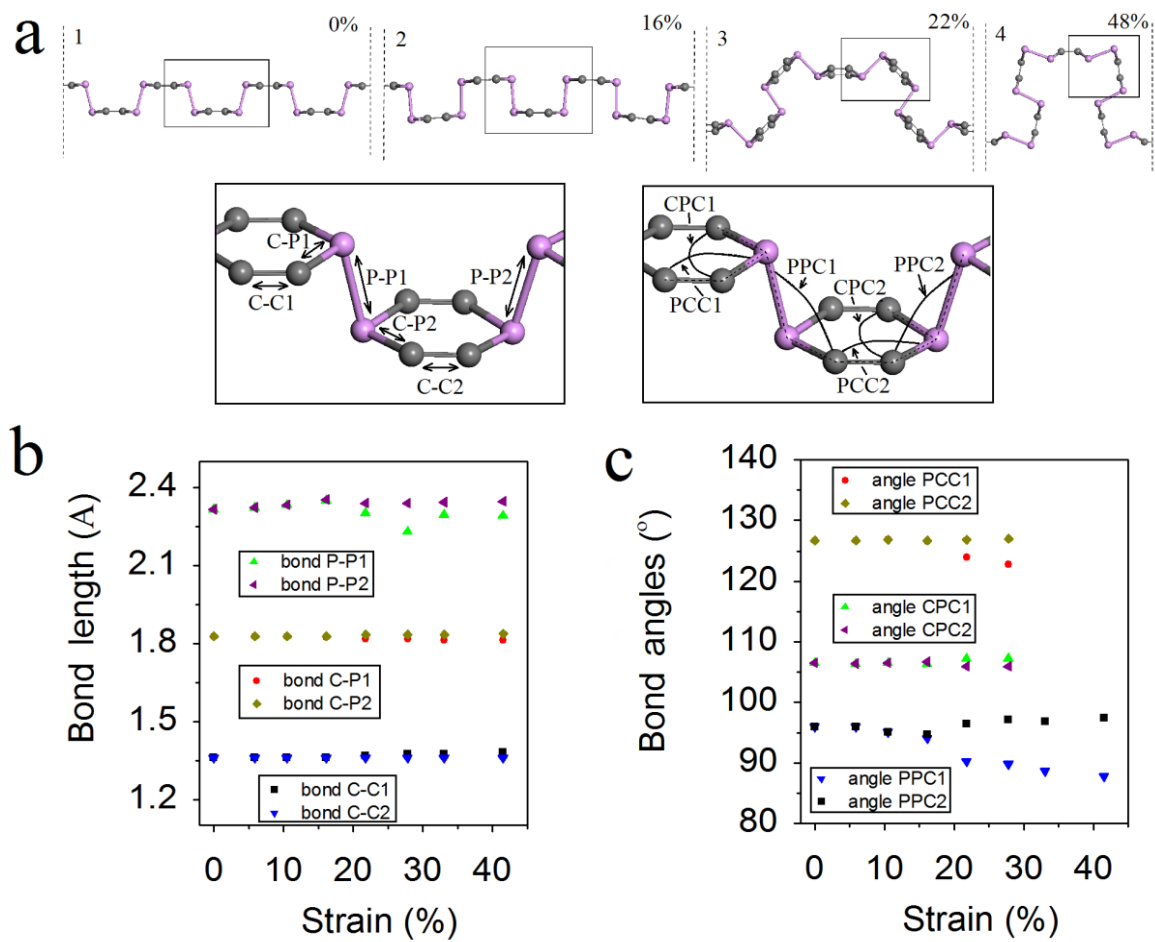

**Figure 3.** (a) A schematic transformation of α-PC under increasing compressive strain (upper panel). The definition of the structural parameters for α-PC (lower panel). The black squares in upper panel show the area where the structural parameters are measured. The variation of (b) bond lengths and (c) hinge angles of α-PC as a function of compressive strain.

**3.3. Atomic structure, electronic and mechanical properties of α-PC nanotubes.** Since the formation of α-PC nanotubes by the rippling of α-PC monolayer under large compressive strains was found to be possible, the atomic configurations and electronic and mechanical properties of these nanotubes were investigated. To determine the smallest size of a stable α-PC nanotube (PCNT), AIMD simulations were conducted for PCNTs consisting of 8 (PCNT8), 16 (PCNT16), 24 (PCNT24), 32 (PCNT32), and 40 (PCNT40) atoms. Movie 2 in SI shows the AIMD results for PCNT40, while its atomic structure is shown in Fig. 4a. Meanwhile, the atomic structures of PCNT24 and PCNT32 are shown in Fig. S2a. According to our calculations, PCNT24, PCNT32, and PCNT40 are direct band gap semiconductors with band gap sizes of 0.29, 0.57, and 0.67 eV, respectively. The Young's moduli of PCNT24,

PCNT32, and PCNT40 along their nanotube axes are 315.87, 328.20, and 333.92 GPa. On the other hand, their Poisson's ratios are 0.20, 0.23, and 0.28, respectively. The calculated cohesive energies of PCNT24, PCNT32, and PCNT40, which are -6.695, -6.741, and -6.743 eV/atom, respectively, suggest that the stability of the nanotube slightly increases with its size. Our prediction also suggests that the PCNTs have a Young's modulus about three times less than that of carbon nanotubes[82,83] and while about two times higher than that of phosphorene nanotubes.[84,85] Meanwhile, the Poisson ratio of the PCNTs is comparable to that of carbon nanotubes.[86]

Fig. 4b presents the band gap size of PCNT40 as a function of the strain, both compressive and tensile, applied along its nanotube axis (Fig. 4a, upper panel). Here, we shall only discuss the results for PCNT40 since PCNT24 and PCNT32 display similar behaviours (Fig. S2 and SI for more details). The band gap of PCNT40 decreases rapidly from 0.67 to 0 eV as the compressive strain increases from 0 to 9%. Meanwhile, at a compressive strain of 3%, the conduction band minimum shifts from between the F and Q points to a Γ point (Fig. S3), signifying a direct-to-indirect band gap transition. Furthermore, at a compressive strain of 9% and higher, the band gap disappears completely, affirming a semiconductor-to-metal transition. An increase in the tensile strain from 0 to 9% leads to an increase of the band gap size from 0.67 to 0.95 eV (Fig. 4b). A rapid decrease of the band gap size is observed when the tensile strain is increased from 9 to 20%. Increasing the tensile strain from 20 to 30% results in the termination of the band gap and a remarkable structure deformation. In particular, a drastic decrease of PCC1 and PCC2 angles, an increase of CPC1 and CPC2 angles, as well as a significant elongation of CP1 and CP2 bonds are observed. Further increasing of a tensile strain would cause nanotube to break.

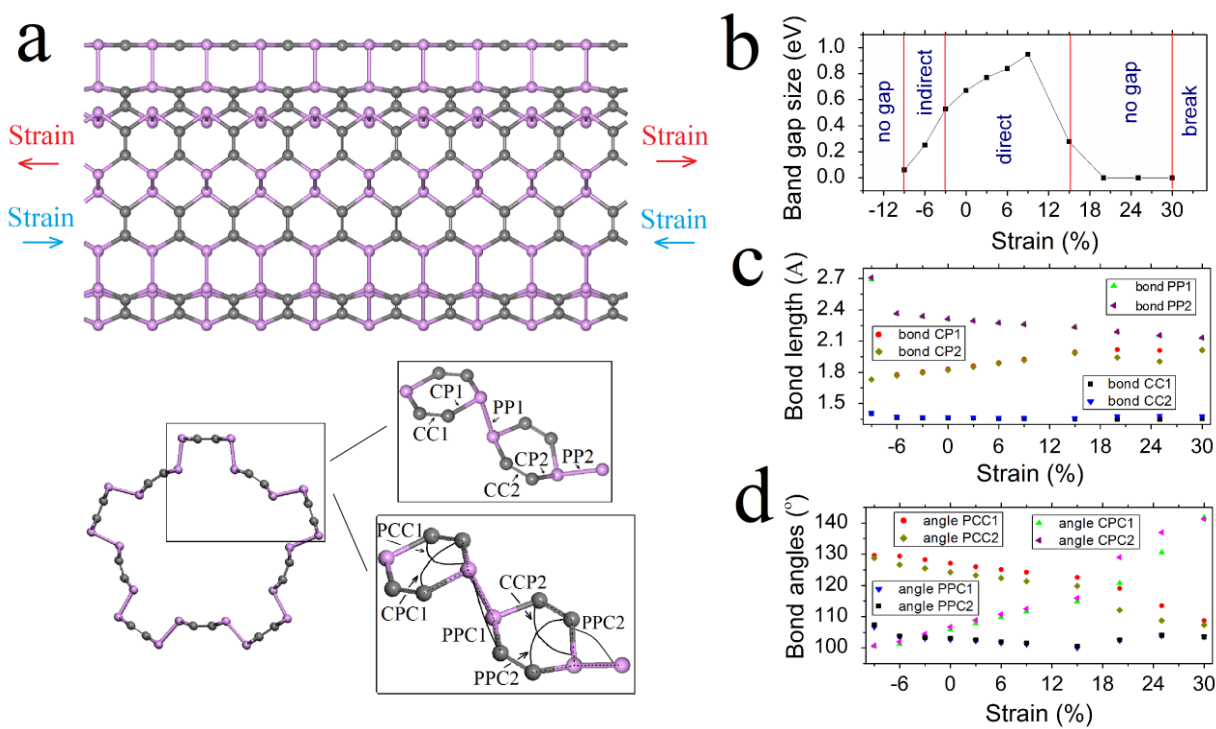

**Figure 4.** (a) Upper panel: a schematic structure of PCNT40 subjected to strain along the its axis. Lower panel: an illustration of the structural parameters of PCNT40. The black square shows the area where parameters are measured. (b) The band gap size of PCNT40, as a function of strain. The bond (c) lengths and (d) angles of PCNT40, as functions of strain.

To understand the changes in the band structure of PCNT40 induced by strain, its structural parameters were analyzed. The following structural parameters (Fig. 4a, lower panel) were tracked: i) the bond lengths PP1 and PP2 connecting the hinges, ii) the bond lengths of the hinges CC1, CC2, CP1, and CP2, and iii) the hinge angles CPC1, CPC2, PCC1, PCC2, PPC1, and PPC2. The variation of the bond lengths and angles are presented in Figs. 4c and d, respectively. As shown, the band gap size correlates positively with the bond lengths and angles. Under a compressive strain increasing from 0 to 6%, the band gap size slowly decreases. In addition, the following minor changes under these strains are observed: the bonds PP1 and PP2 are slightly elongated and the bonds CC1 and CC2 remain almost unchanged, while the bond angles PCC1, PCC2, PPC1, and PPC2 linearly increase. When the compressive strain is increased from 6 to 9%, the following significant changes in the structural parameters have taken place: the lengths of bonds PP1 and PP2 drastically increase and bonds CC1 and CC2 are also elongated noticeably, while the PCC1 angle decreases and the PPC1 and PPC2 angles significantly increase. In addition, there is an almost linear positive correlation between the tensile strain and the band gap size, the bond lengths CP1 and CP2, and the bond angles CPC1 and CPC2. On the other hand, the tensile strain exhibits a negative linear correlation with the bond lengths PP1, PP2, CC1 and CC2, as well as the bond angles PCC1, PCC2, PPC1, and PPC2. It can be concluded that direct-indirect and semiconductor-to-metal transitions in PCNT40 occur mainly due to a transformation of the bond lengths and angles.

To further understand the effects of strain on the electronic structure of PCNT40, we calculated the electron localization function (ELF) for PCNT40 under different strains. For PCNT40 under the compressive strain of 9%, and PCNT40 under the tensile strain of 20% the isosurface value of 0.65 is adopted in Figs. 5a-c. For the same structures in Figs. 5d-f, the value of the ELF maps (between 0 and 1) reflects the degree of charge localization in the real space, where 0 represents a free electronic state while 1 represents a perfect localization. As shown in Figs. 5b and e, the electrons are packed together at C-P bonds (the red circle) and slightly delocalized at P-P (the black circle) bonds. As indicated in Figs. 5a and d, the compression strain of 9% leads to an increase of electrons at P atoms and its localization at C-P bonds (the red circle) while at P-P bonds (the black circle) electrons are almost delocalized. That suggests the nearly broken P-P bonds. At the tensile strain of 20% (Figs. 5c and f) the electrons at P-P bonds exhibit a more highly concentrated distribution while at C-P bonds their localization decreases. That indicates a strong covalent chemical bonding between P atoms.

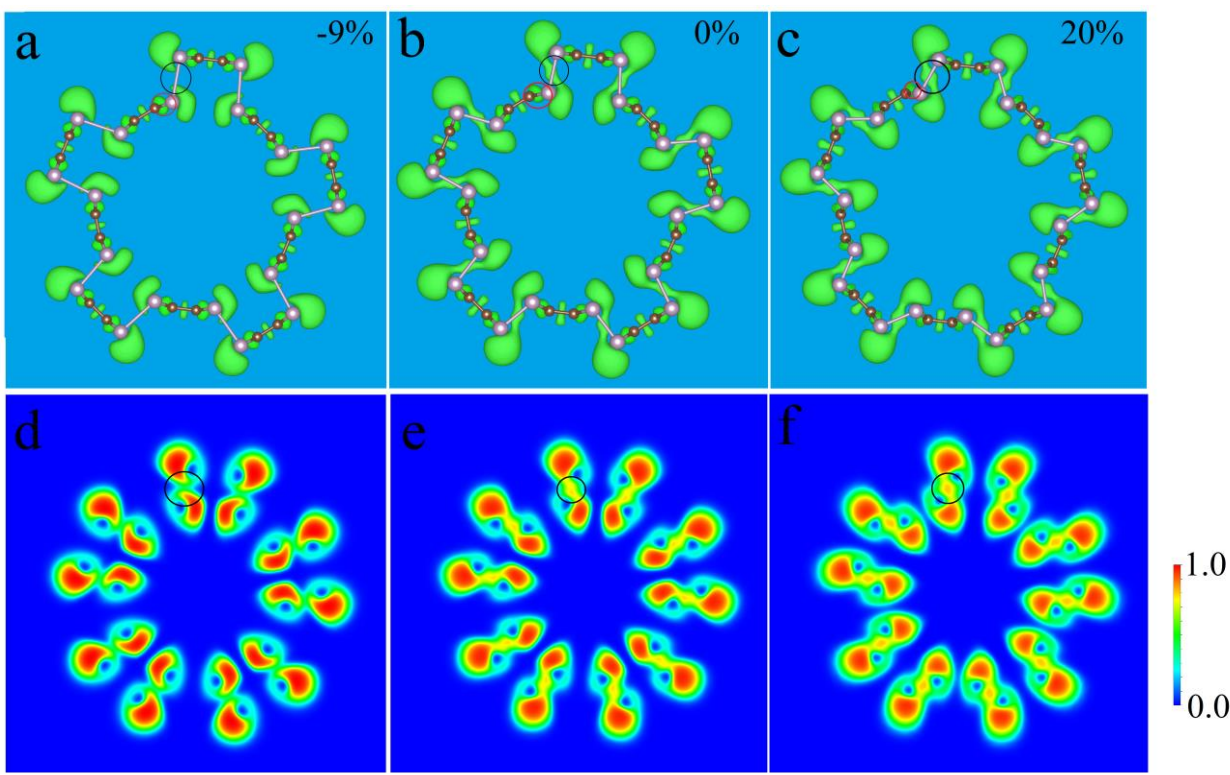

Figure 5. The ELFs (upper panel) and ELF maps (lower panel) for PCNT40 under (a) compressive strain of 9%, (b) unstrained, and (c) under tensile strain of 20%. The value for ELFs in (a)-(c) is set to 0.65.

## 4. CONCLUSIONS

By applying first-principles simulations, we examined and explained the modification of the electronic and mechanical properties of $\alpha$-PC upon rippling under a large compressive strain along the armchair direction. It was found that the Young's modulus of rippled $\alpha$-PC varies from 290.41 to 348.69 GPa under different compressive strains. Rippling also causes an indirect-direct band gap transition in $\alpha$-PC and modifies its workfunction. In addition, the possibility of restructuring of $\alpha$-PC into a nanotube-like structure under extreme compressive strains was revealed. $\alpha$-PC nanotubes of different sizes were subsequently simulated and studied, and the smallest diameter for an $\alpha$-PC nanotube to be stable at 300 K were evaluated. The electronic and mechanical properties of stable $\alpha$-PC nanotubes of different diameters were also investigated. The Poisson's ratio of $\alpha$-PC nanotubes was not inferior to that of phosphorene nanotubes, while their Young's modulus was found to exceed that of phosphorene nanotubes. The investigated $\alpha$-PC nanotubes possess a highly tunable band gap under compressive and tensile strains applied along the nanotube axes. Moreover, we showed that through strain engineering, it is possible to achieve a semiconductor-to-metal transition in $\alpha$-PC nanotubes, or significantly enlarge their band gap size. Therefore, our work suggests rippling as an effective method to tune the mechanical and electronic properties of $\alpha$-PC monolayer. We also show the possibility of PCNTs fabrication by rippling of a $\alpha$-PC monolayer under extreme compressive strains. Furthermore, due to the new-found extraordinary properties of PCNTs, they may eventually replace phosphorene and graphene nanotubes in the designing of nano-devices, and find application in straintronic, optical and photovoltaic devices.

## ASSOCIATED CONTENT

The Supporting Information (SI)

Electronic structure of the α-PC monolayer under compressive strains, atomic and electronic structures of PCNT24 and PCNT32, the band structures of PCNT40 under compressive strains (PDF), AIMD results of transformation of α-PC from the rippled structure into the nanotube-like structure (Movie 1), AIMD results on structural stability of PCNT40 (Movie 2).

## AUTHOR INFORMATION

**Corresponding Authors**

*E-mail: andrey.kistanov@oulu.fi

## ACKNOWLEDGMENTS

The authors wish to acknowledge CSC – IT Center for Science, Finland, for computational resources. E.A. Korznikova thanks the Russian Foundation for Basic Research, grant No. 18-32-20158 mol_a_ved. K. Zhou acknowledges the financial support provided by Nanyang Environment and Water Research Institute (Core Fund), Nanyang Technological University, Singapore. A.A. Kistanov acknowledges the financial support provided by the Academy of Finland (grant No. 311934).

# Supplementary Information

## 2D Black Phosphorus Carbide: Rippling and Formation of Nanotubes

*Stepan A. Shcherbinin[†], Kun Zhou[‡], Sergey V. Dmitriev[§,∥], Elena A. Korznikova[§], Artur R. Davletshin[⊥], \*Andrey A. Kistanov[‡]*

[†]Southern Federal University, 344006 Rostov-on-Don, Russia

[‡]School of Mechanical and Aerospace Engineering, Nanyang Technological University, 639798 Singapore

[§]Institute for Metals Superplasticity Problems, Russian Academy of Science, 450001 Ufa, Russia

[∥]National Research Tomsk State University, 634050 Tomsk, Russia

[⊥]Department of Petroleum and Geosystems Engineering, The University of Texas at Austin, Austin 78712, TX, USA

[¦]Nano and Molecular Systems Research Unit, University of Oulu, 90014 Oulu, Finland

**\*Email**: andrey.kistanov@oulu.fi

# Electronic structure of the *α*-PC monolayer

**Figure S1.** The atomic (upper panel) and the band (lower panel) structures of rippled *α*-PC monolayer under compressive strain in a range from 0 to 48%.

## Atomic and electronic structures of α-PC nanotube

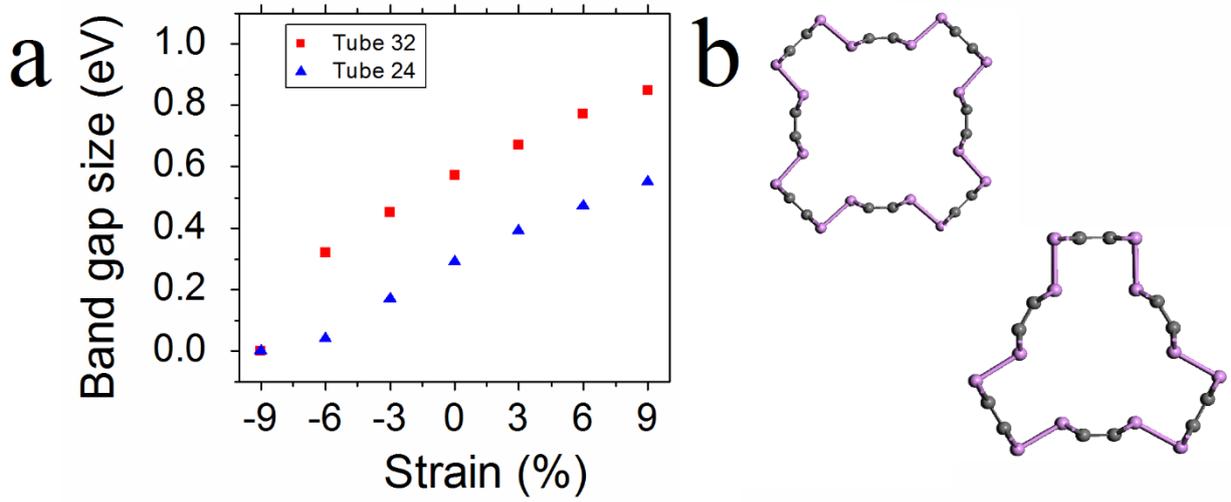

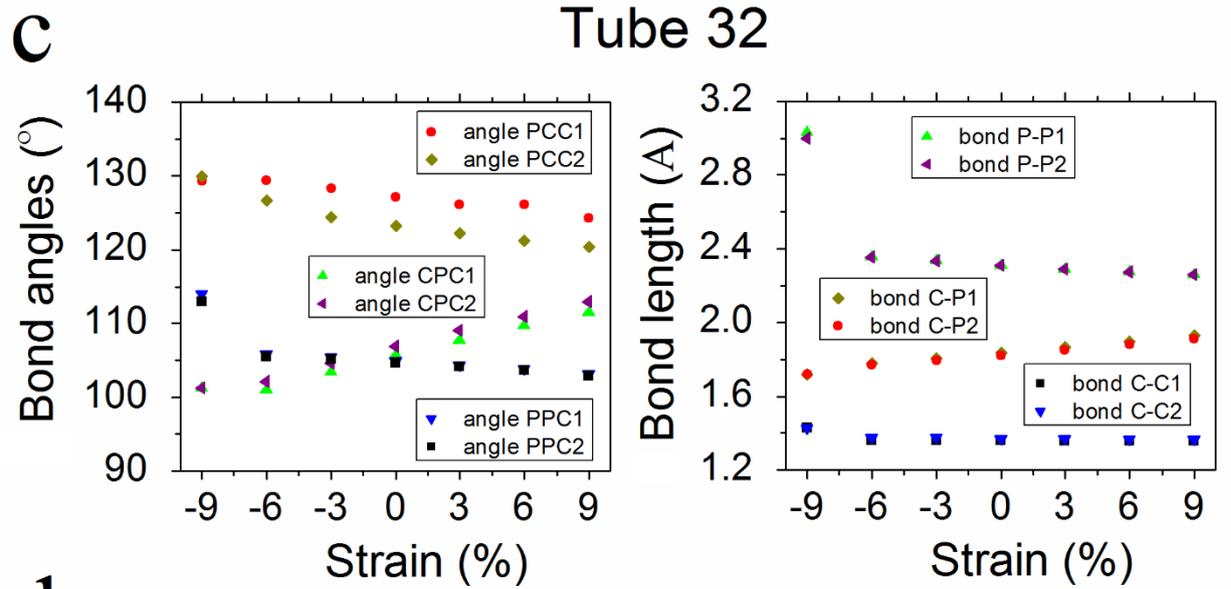

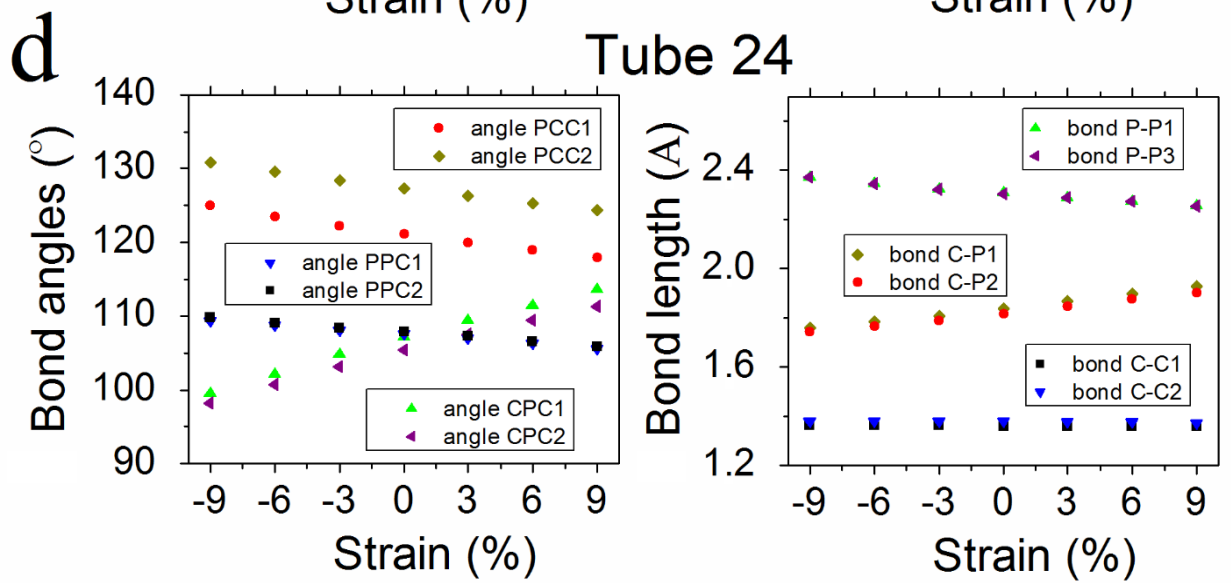

**Figure S2.** (a) The band gap size of PCNT24 and PCNT32, as a function of strain. (b) The structure of PCNT24 and PCNT32. The variation of bond lengths and angles of (c) PCNT24 and (d) PCNT32, as a function of strain.

Figure S2a presents the band gap size of PCNT24 and PCNT32 as a function of strain applied along the nanotube profile. The PCNT24 and PCNT32 show similar behaviours as PCNT40. The band gap of PCNT24 (PCNT32) decreases rapidly from 0.29 (0.57) to 0 eV as the compressive strain increases from 0 to 9%. A tensile strain ranging from 0 to 9% leads to an increase of the band gap size of PCNT24 (PCNT32) from 0.29 (0.57) to 0.55 (0.85) eV.

To understand the changes in the band structure of PCNT24 and PCNT32 (Figure S2b) induced by strain their structural parameters were analyzed similarly to the case of PCNT40. The following structural parameters are tracked: i) the bonds' lengths P-P1 and P-P2 connecting the hinges; ii) the bonds' lengths of the hinges C-C1, C-C2, C-P1, and C-P2; iii) the hinge angles CPC1, CPC2, PCC1, PCC2, PPC1, and PPC2. The variation of the bond lengths and angles for PCNT32 and PCNT24 are presented in Figure S2c and d, respectively. As it is seen, the change of band gap size is correlated with the changes in the bonds' lengths and angles. Under compression, the band gap size slowly decreases with increasing a compressive strain from 0 to 6 %. Similar minor changes at these strains were observed in dependencies: the bonds P-P1 and P-P2 slightly elongates and the bonds C-C1 and C-C2 remain almost unchanged, the bonds' angles PCC1, PCC2, PPC1, and PPC2 monotonically increase. At the compressive strain of 9%, when the band gap size drops down, the following significant changes in the structural parameters take place: the lengths of bonds P-P1 and P-P2 drastically increase and bonds C-C1 and C-C2 also elongate noticeably, the PPC1, PPC2, PPC1, and PPC2 angles increases and the CPC1 and CPC2 angles significantly decrease. In addition, almost linear increase of the band gap size upon increasing of a tensile strain correlates with linear increase of the bond lengths and angles.

**Electronic structure of *α*-PC nanotube under a tensile/compressive strain**

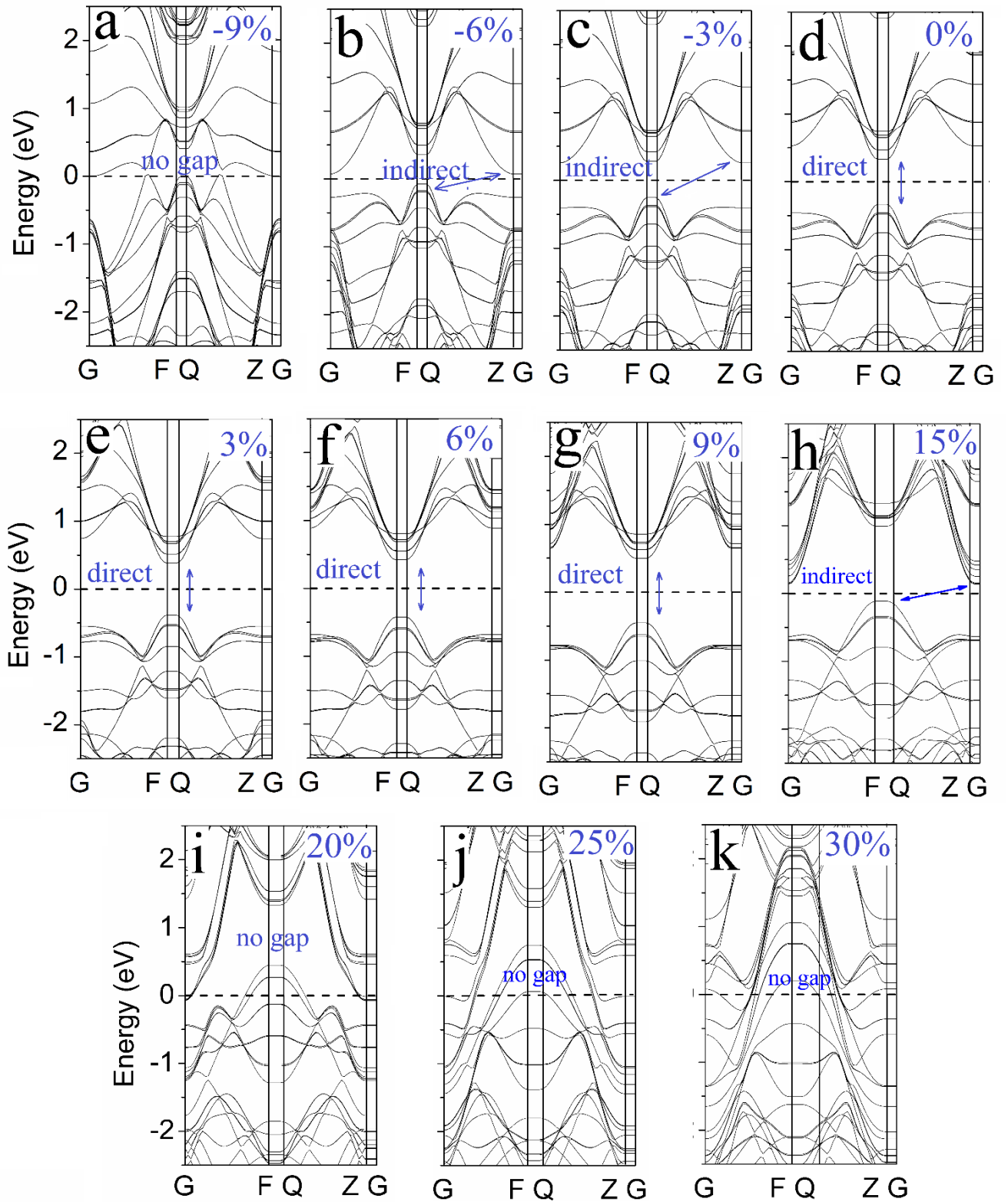

**Figure S3.** The band structures of PCNT40 under compressive strains of (a) 0%, (b) 3%, (c) 6%, and (d) 9% and tensile strains of (e) 3% (f) 6%, (g) 9%, (h) 15%, (i) 20%, (j) 25%, and (k) 30%.